# Thinking fast and slow - decision intelligence for power systems

Apoorv Mathur
Siemens Grid Software, IEEE Senior Member,
Seattle, Washington, US 98109
apoorv.mathur@ieee.org

*Abstract*— **Decision-making in power systems spans multiple timescales—from milliseconds to prevent surges, to seconds to balance frequency and protect grid assets, to minutes for real-time energy balancing, to day-ahead, seasonal, and long-term planning. Growing uncertainty and complexity, driven by intermittent renewables and distributed energy resources (DER), demand fresh approaches to power system intelligence and architecture. Daniel Kahneman describes the interplay of two systems of human decision-making: System 1 (fast, intuitive, experience based, reactive), and System 2 (slow, deliberate, analytical). Similarly, octopus intelligence illustrates a model for distributed yet coordinated decision-making between central and edge intelligence. Future power systems must embed coordinated intelligence that operates across diverse timescales and with placement at both edge and centralized levels. This paper maps decision-intelligence in power systems against System 1 and 2 and edge-central architecture paradigms based on the trade-offs inherent in decision making such as speed/latency, energy cost/compute, accuracy, and robustness. The framework inspires an agentic intelligence architecture – laying the foundation for trustworthy, autonomous power systems of the future.**

## I. Introduction

Cognitive science provides a basis for human decision making under ambiguity, complexity and uncertainty. Cognitive processes for rational decision making range from perception, attention, learning, mental modeling, prediction, critical thinking, reasoning, goal setting, planning, control and adaptation. One of the seminal works in cognitive science is Thinking, Fast and Slow, by Nobel laureate Daniel Kahneman [1]. The work demonstrates the interplay of two decision making mechanisms: System 1, which is fast, intuitive, and reactive, and System 2, which is slow, deliberate, and analytical. Human behavior and intelligence operates at multiple time-scales and layers – as demonstrated in cognitive and behavioral science literature – with feedback loops across the layers [2]. Another evolutionary model of intelligence is the Octopus – which enables distributed yet coordinated intelligent decisions [3]. The architectures are guided by evolutionary design to meet the needs for short term survival and the long term optimization leading to decision making under trade-offs of speed, cost (energy use), accuracy and robustness of required decisions.

Decision making in large scale power systems is a widely researched topic – decision making ranges in timescales from the microsecond in operations to the decades long planning cycle – as well as – from an individual edge device to a micro-grid, feeder or sub-circuit to the distribution and transmission networks. A wide variety of information, algorithms and technology architectures support decision making in this system – some decisions human-in-the-loop, while others are pre-programmed or autonomous. Decision intelligence techniques applied in power systems range from rule based, heuristic, neural nets, simulation, probabilistic to optimization. Architecture of decision-making systems span centralized, modular, distributed at the edge, hierarchical. Finally, with the advent of Agentic AI, we can support the orchestration across multiple types and scales of intelligence.

While power systems have grown to manage ambiguity, complexity and uncertainty over a period of time, we look through the evolutionary needs lens for decision intelligence, and make parallels with systems of human cognition to derive insights on how power systems of the future would evolve.

The rest of this paper is organized in the following sections: II reviews similar themes explored in past research, III outlines the decision making processes along with the guiding considerations on the trade-offs of speed, cost (energy use), accuracy and robustness, IV covers systems coordination, V maps the decision making processes in power systems with the framework, VI draws inspiration from the framework into a modern software agentic architecture, VII summarizes contributions and outlines direction of future work.

## II. Past Work

There is a lot of literature on the techniques and architecture for decision intelligence across power systems planning and operations. Key architectural concepts investigated digital twins [4, 5], hierarchical, distributed structures [6, 7] and autonomous energy systems [8]. Key techniques range across simulation, optimization- central and distributed [10-15] with emerging emphasis on leveraging machine learning, neural networks and artificial intelligence over the last decade [16-25]. However, there's limited investigation of the based on cognitive inspired designs for decision making [9].

## III. Decision Making Frameworks

Decision making is linked to goals and follows a sense-think-act cycle given a goal. An extension to this is the perception, state representation, prediction, planning (includes generation, evaluation, and selection of alternative), action, feedback and adaptation cycle.

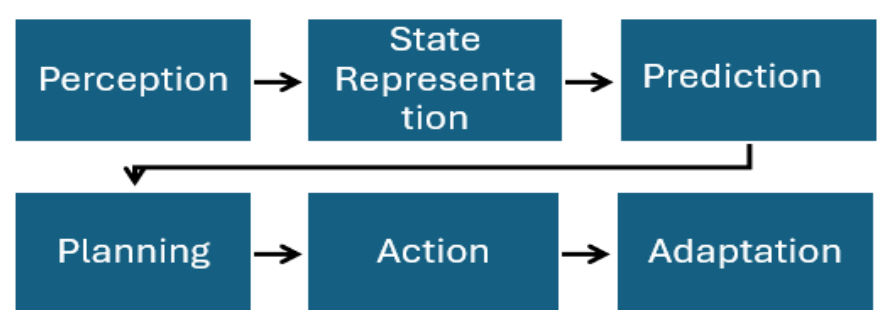

Fig. 1. Decision making cycle

**Systems 1 and 2**

Humans have evolved to survive and achieve their goals in an energy efficient manner. Decisions that impact survival often need to be quick and, can disregard short-term-cost – these often require quick actions based on System 1 thinking. In contrast, decisions that have longer term consequences, have a high degree of uncertainty, ambiguity or uncertainty require explicit planning and analytical thinking with higher accuracy and optimization often rely on System 2. As an example, driving on a daily route is often automatic System 1 activity; whereas a new destination leads System 1 to lower than threshold certainty and accuracy triggering activation of System 2 – requiring it to pay attention to plan route and turns explicitly. However, context provided by System 1 to System 2 about the specific detour and options enables System 2 to narrow focus.

Governing factors on System choice are impact/objective (survival vs optimization), availability of experience and a tradeoff between speed (time available), cost (energy use), quality - *accuracy, robustness and interpretability*♣ (is learned policy applicable across scenarios, interpretable and adaptable).

TABLE I. FACTORS IMPACTING THINKING MODE FOR DECISION MAKING

| **Mode** | **Impact** | **Experience** | **Speed** (latency) | **Cost** | **Quality**♣ |
|---|---|---|---|---|---|
| System 1 | Survival | Yes | Fast | Low | Low |
| System 2 | Optimization | No | Slow | High | High |

**Humans and the Octopus**

Back to the human nervous system as a reference point for decision intelligence architecture – we possess a reflex system – meant for quick response based on localized sensing and a central brain that synthesizes all sensory inputs across all senses and can pair it with long term goals to make a decision.

An octopus takes this one step further toward a hierarchical architecture with greater local sensing and decision making – that is orchestrated centrally. This is the case with each of the octopus' arms and their coordination with its central brain. The octopus arms have nearly double the neurons compared with the central brain. Octopus arms can make decisions based on local sensing such as 'tastes like food' triggering action 'grab' while coordinating the goal – say 'catch prey' with the central brain.

Governing factors for decisions to be closer to the edge or central are: impact/objective (local or system wide), and a tradeoff between speed/ latency (time available), operations cost (energy on communication, data processing, computation), data access♦ (data availability, local/global use, privacy, communication latency), quality - *accuracy and robustness*♠ (noise, reliability of signal and computation).

TABLE II. FACTORS IMPACTING ARCHITECTURE FOR DECISION MAKING

| **Placement** | **Impact** | **Data Access**♦ | **Speed** (latency) | **Cost** | **Quality**♠ |
|---|---|---|---|---|---|
| Edge | Local | Local | Fast | Low | High |
| Hierarchical | System | Med | Slow | High | Med |
| Central | System | Global | Slower | High | Low |

## IV. SYSTEM OF SYSTEMS COORDINATION

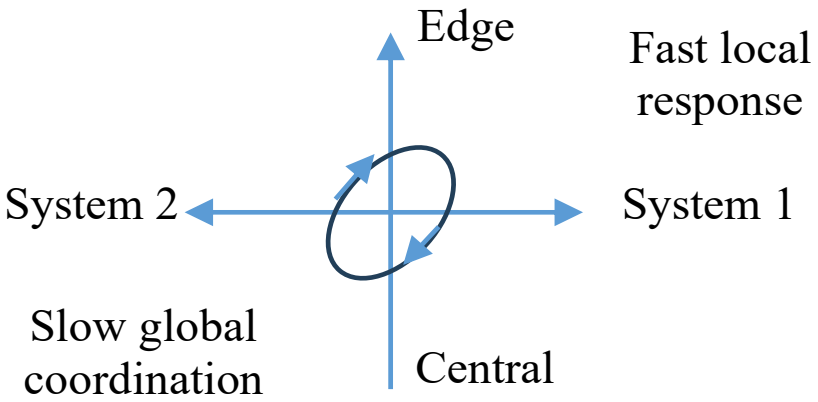

Fig. 2. Coordinated systems of thinking and edge-central

Although System 1 and System 2 function differently, they typically work together rather than independently. System 1 operates continuously, drawing on experience, operating within tight response times and costs, and engages System 2 when faced with uncertainty, complexity, or ambiguity. The feedback loop from System 2 corrects System 1's errors. Similarly, in case of the octopus, the coordination loop extends further to edge and central intelligence – where edge can produce locally optimal outcomes but coordinate with central goals over time.

## V. MAPPING POWER SYSTEM DECISIONS

Decision making in grid planning and operations happens across multiple timescales. Here too, there is some similarity in terms of the objective of the system in the very short run is to be reliable, to avoid failure, or damage to the system. In the medium and long term, decision making is analytical for efficient decision making while satisfying the goals and constraints.

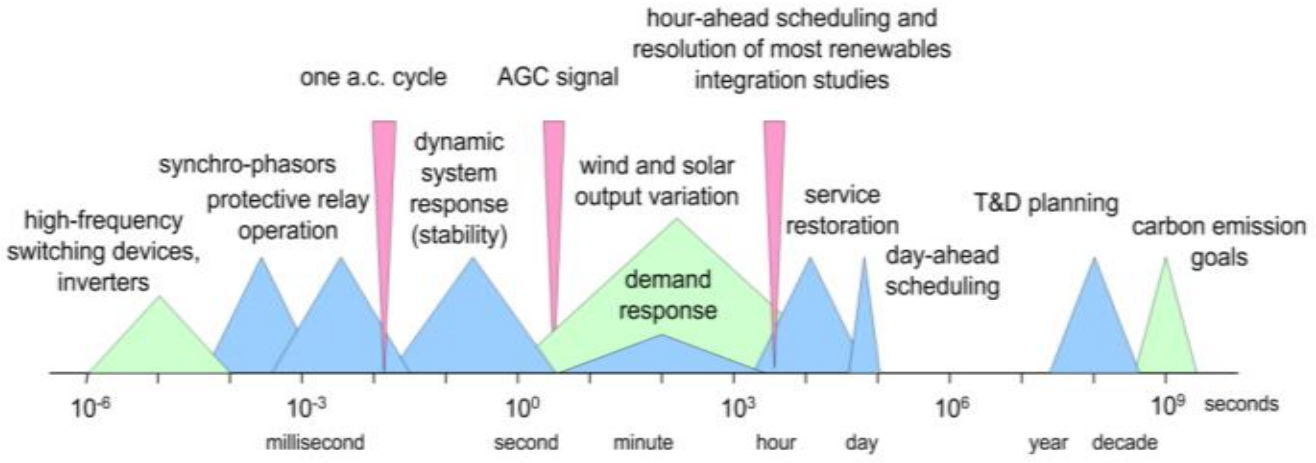

Fig. 3. Timescales in Electric Grid Planning and Operations

TABLE III maps power system decisions into System 1 or 2, as well as, into central or edge decisions based on factors above.

One example of decisions made in milli-second (ms) time-frame with local sensing is of grid protection. Protective relay devices are designed to trip a circuit breaker when a grid problem such as an overcurrent occurs. Relays operate within 16–32 ms. This is a reflex actions or System 1 driven action. However, once this local protection action is taken, the central systems need to coordinate with fault location isolation and service restoration (FLISR) to optimize the outage and restoration within a few minutes – akin to a System 2 response.

Managed EV charging constrained by distribution grid limits - represents local control guided by global grid constraints—akin to an octopus arm moving independently while remaining coordinated with the body's intent. EV charger dynamically adjusts its charging rate based on signals such as feeder capacity, transformer load, or local voltage conditions. These limits would be established centrally by the utility, but executed autonomously at the charger using edge intelligence.

TABLE III. DECISION MAP TO SYSTEM TYPE – SYSTEM 1 OR SYSTEM 2 AND CENTRAL OR EDGE INTELLIGENCE

| Decision | Time | Mode: System 1 or 2 | Placement: Central or Edge | Description | Key Considerations |
|---|---|---|---|---|---|
| Grid Protection (e.g., relay tripping) | 16–32 ms | System 1 (Reflex) | Edge (relay device) | Relay detects overcurrent, trips circuit breaker in ms | Speed – protect equipment, ensure high recall |
| Fault Isolation & System Restoration (FLISR) | Mins | System 2 (Analytical) | Central coordination | Automatic system isolation and restoration following a fault | Speed – restore quickly, accurately to maintain reliability |
| Automatic Generation Control | Secs | System 1 (Reflex) | Edge (generator) | Automatic response to maintain frequency | Speed – ensure reliability |
| Dispatch Decisions | Mins | System 2 (Analytical) | Central orchestration | Day-Ahead, Real-Time energy market balance-optimization | Optimization, high complexity to maintain reliability |
| Peak Load Shifting (e.g., smart EV charging) | Hours | System 2 (Analytical) | Central (global signals) & Edge (local sensing) | Manage EV charger schedules based on time-of-use price or dynamic grid capacity signals | Optimization schedule globally coordinated |
| Long-Term Grid Planning | Days | System 2 (Analytical) | Central | Resource allocation and infrastructure planning | Decision under high uncertainty, strategic objectives |

As a corollary to human decision intelligence, power systems use a host of algorithmic techniques offering a spectrum of decision-making intelligence ranging from rules, power flow, simulation, optimization to machine learning, neural networks, deep learning and reinforcement learning. Most literature doesn't map the type of decision system; however, based on the factors above, we classify heuristics, rules, anomaly detection, neural networks, reinforcement learning relate to fast, intuitive, experience-based System 1; whereas, analytical, power flow, simulation, optimization to the analytical System 2.

Even though hardware and software capabilities make complex computation feasible in a short time span, given the large scale of the systems, latency needs still make the systems thinking and central-edge architecture decisions relevant to decision systems design. With more pervasive, lower cost computational hardware at the edge such as AMI 2.0 smart meters, inverters, near real time bi-directional communication network, improving network availability, smarter centrally coordinated distributed decision making is increasing [26, 27].

## VI. ARCHITECTURE FRAMEWORK

Today, agentic artificial intelligence (*Agentic AI*) can autonomously operate based on goals, can plan, reason and operate across time-horizons. Agents can operate based on the same base set of underlying capabilities, data sets to service the goals across roles and time-horizons.

With this new paradigm, there is an opportunity to redesign software systems beyond traditional application centric architectures that dominate the industry (*MDM, GIS, Simulation, ADMS, DERMS*), which operate within data silos, are relevant only for particular roles and time-scales (planning, operations, customer programs, asset health etc.) [29-32].

We envision goal based agentic decision systems for power systems based on the systems of thinking and central – edge agent coordination frameworks described above – that can span across data-silos, roles and timescales. Overall, the previously described coordinated approach between System 1 and System 2 – along with centrally coordinated edge behavior will be critical to achieve trustworthy decision intelligence to manage power systems of the future.

In our proposed framework, an orchestrator agent facilitates decision making across Systems 1 and 2 agents based on the goal/impact scope, data availability, response time, compute resource availability (cost), required accuracy and robustness under the uncertainty, complexity, and ambiguity of the situation. In this way, multiple agents with different algorithms and compute architecture can support the goals of the system within time and resources available - mirroring bounded optimality in AI literature [28].

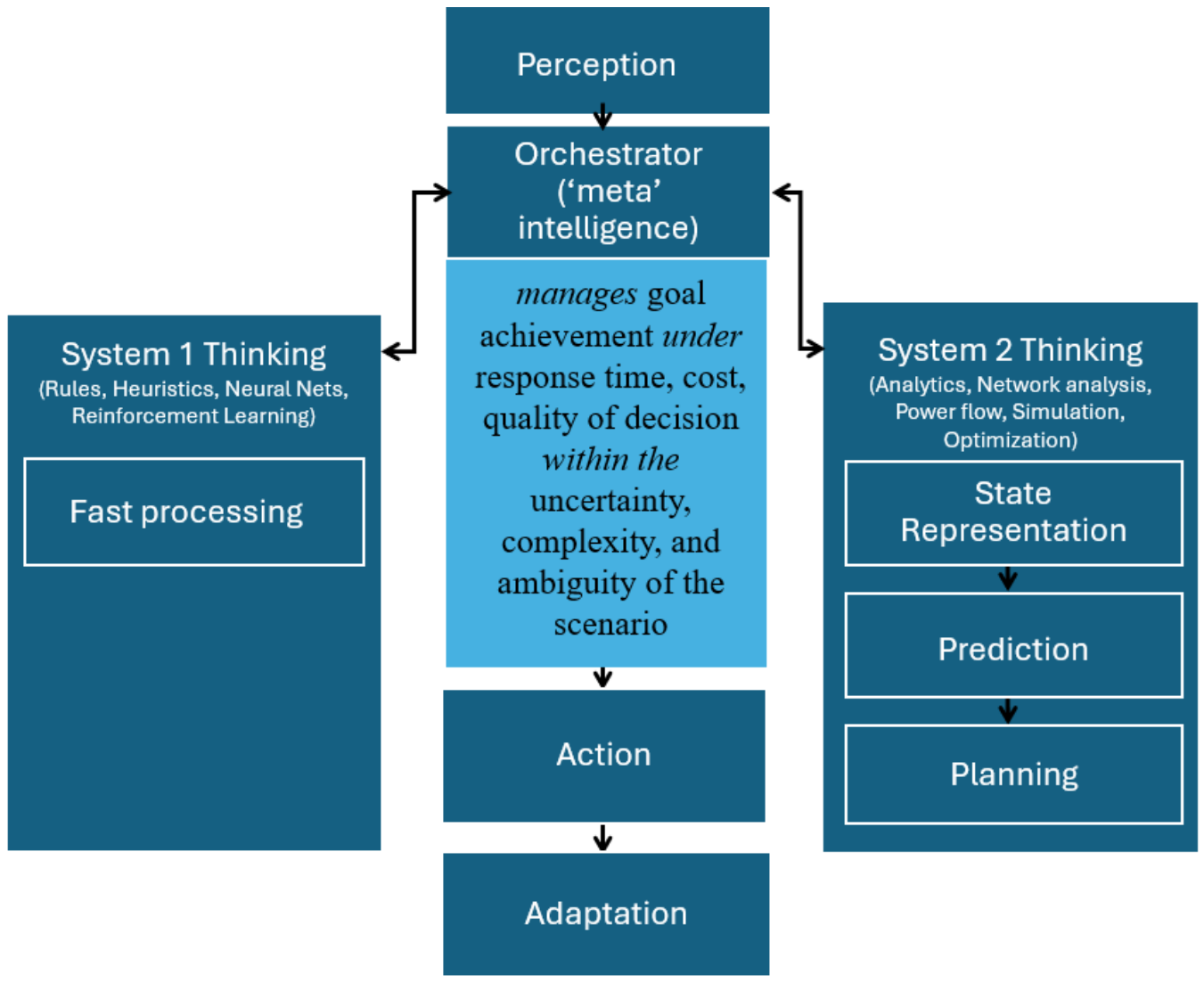

Fig. 4. Coordinated systems of thinking in decision making (extend Fig 1)

The coordination loop can similarly be extended to hierarchies of central-edge agent units that coordinate to achieve the goals of the system as well as individual edge units.

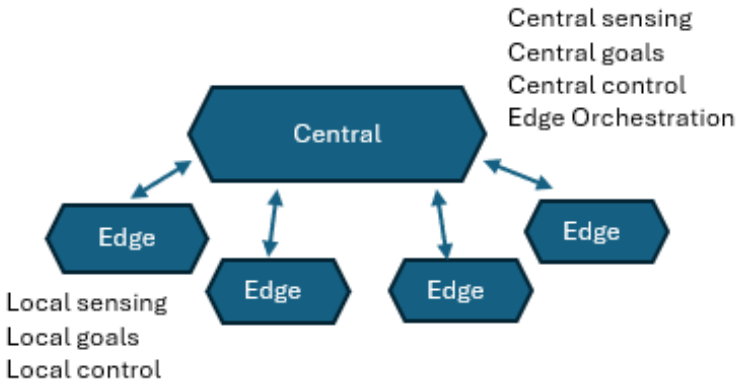

Fig. 5. Coordination between edge and central decision making

**Examples of goal based agents that operate across timescales**

Below are illustrative examples structured as goal oriented agents that combine System 1 and System 2 thinking, leveraging edge-central coordination:

**Fault detection and self-healing agent:**

As described in the previous section, consider an agent that has a goal of minimizing fault occurrence and impact. This autonomous fault detection & self-healing AI agent monitors grid conditions, detect faults triggered by reflexive Relay responses (System 1) whereas, a central agent autonomously optimizes the fastest and least power loss approach to reroute power, isolate sections and restore power (System 2).

**Power quality management and loss minimization agent:**

When a power quality deviation is perceived at a feeder level, the orchestrator agent responsible for managing the power quality while ensuring system efficiency triggers two actions. An edge agent applies pre-coordinated volt-watt logic to adjust inverter setpoints to avoid violations in real-time with local sensing (System 1); while the a second central agent responds on a slower cycle, optimizing the entire grid for efficiency using optimal power flow to set tap changer, voltage regulator and capacitor bank settings (System 2).

**Contingency planning agent:**

Preparing for grid contingencies involves N-1 scenario analysis. A simple N-1 contingency scenario for a critical grid component is too narrow, whereas a Montecarlo approach that tries to account for a average scenario over a number of such N-1 events is highly computationally intensive. In such a case, an agent may choose a machine learning based approach that leverages past contingency scenarios (System 1) to help narrow down focus on realistic scenarios for grid analytics (System 2) balancing speed and accuracy without exhaustive computation.

These examples demonstrate how System 1 provides rapid, experience-based actions and filters complexity, while System 2 delivers deeper optimization and planning. Together, they enable orchestrated intelligence across edge and central layers, ensuring both responsiveness and robustness.

## VII. Summary, Conclusions and Future Work

In this paper, we describe a goal based framework for decision making in power systems that is inspired by human systems of thinking 1 and 2 as well as on the central and edge systems based on the octopus – and the coordination of these systems. We describe the tradeoffs that help us understand when each of these systems should be used, as well as, how these systems work in tandem.

We then describe how decisions in power systems map to this framework and extend the framework to guide the architecture of decisions intelligence system for power systems. We conclude by connecting the framework with the Agentic AI approach for modern software – and provide examples of intelligent agents that leverage the systems thinking approach to make autonomous decisions spanning across timescales.

We expect that the framework will enable demonstration of use cases of goal based decision agents for power systems that are able to orchestrate and navigate across the systems 1 and 2, coordinate with edge systems. Our hope is that the new approach helps move to toward autonomous power systems that operate across silos, teams and timescales while managing the response times, as well as the uncertainty, complexity and ambiguity required for the decisions.

To operationalize such an agentic architecture framework for decision intelligence – one needs enabling platform, algorithmic, hardware/software, data and communication capabilities. For our future work, we intend to implement demonstration cases that will highlight practical architectural, platform, algorithmic, data and communication considerations; while demonstrating feasibility, benefits and challenges of pursuing this architecture framework in practice. Additionally, we intend to further investigate the role of the underlying enablers for such decision such as the grid digital twin that acts as the base for the state of the grid – on top of which the described agents can make decisions.

## Acknowledgment

To colleagues Siddharth Bhela, Hakan Özlemiş, Bharadwaj Sathyanarayana Ranganathan, thank you for review and inputs.